\documentclass[letterpaper,11pt]{report}

\pdfoutput=1

\usepackage{graphicx}
\usepackage{latexsym}
\usepackage{makeidx}
\usepackage{url}
\usepackage{fullpage}
\usepackage{hyperref}

\title
{
	Alchymical Mirror: Real-time Interactive Sound- and Simple Motion-Tracking Set of Jitter/Max/MSP Patches
}

{\author
	{\bf
		\\\\
		Elizaveta Eidelman\\
		\url{lizets@yahoo.com}\\\\\\
		Serguei A. Mokhov\\
		\url{mokhov@cse.concordia.ca}\\\\\\\\\\\\
		Concordia University\\
		Montr\'eal, Qu\'ebec, Canada\\
		December 2005
	}
}

\date{Mon Dec 12 12:15:23 EST 2005
}

\makeindex

\newcommand{\file}[1]{\texttt{#1}\index{Files!#1}}

\newcommand{\api}[1]{\texttt{#1}\index{API!#1}}

\newcommand{\lucidL}[1]{{$\mathit{Lucid}$}($L$) }

		{}

\def\myvert{\raise 2.27pt \hbox{\vrule depth 0pt height 8pt width 0.2mm}}
\def\myarrow{\hspace*{0.43mm}%
             \raise 2.29pt\hbox{\vrule depth 0pt height 8pt width 0.16mm}%
             \hspace*{-0.32mm}%
             $\longrightarrow$
             \ %
             }

\begin{document}

	\begin{titlepage}
		\maketitle
	\end{titlepage}

	\begin{abstract}
This document supplements an experimental Jitter / Max/MSP collection of implementation
patches that set its goal to simulate an alchemical process for a person standing
in front of a mirror-like screen while interacting with it. The work involved takes
some patience and has three stages to go through. At the final stage the "alchemist"
in the mirror wearing sharp-colored gloves (for motion tracking) is to extract the
final ultimate shining sparkle (FFT-based visualization) in the nexus of the hands.
The more the hands are apart, the large the sparkle should be. Moving hands around
should make the sparkle follow. To achieve the desired visual effect and the feedback
mechanism, the Jitter lattice-based intensional programming model is used to work
on 4-dimensional (A+R+G+B) video matrices and sound signals in order to apply some
well-known alchemical techniques to the video at real-time to get a mirror effect
and accompanying transmutation and transformation stages of the video based
on the stability of the sound produced for some duration of time in real-time.
There is an accompanying video of the result with the interaction with
the tool and the corresponding programming patches.
	\end{abstract}

	\pagenumbering{roman}
\tableofcontents
\clearpage
\pagenumbering{arabic}


	\chapter{Introduction}
\index{Introduction}

\section{Purpose}
\index{Introduction!Purpose}

This document supplements an experimental
Jitter~\cite{jitter} / Max/MSP~\cite{maxmsp} collection of patches
that set their goal to simulate an alchemical process for a person
standing in front of a mirror-like large LCD screen. The
work involved takes some patience and has three stages
to go through. At the final stage the alchemist
in the
mirror wearing a sharp-colored gloves to extract the final
ultimate shining sparkle in the nexus of
the hands. The
more the hands are apart, the large the sparkle should be.
Moving hands around should make the sparkle follow. This
would be the definition of the
purification process of an alchemist working on a some sort of
Philosopher's Work.

\section{Jitter and Max/MSP}
\index{Introduction!Jitter and Max/MSP}

The Jitter lattice-based intensional programming model is used
to work on 4-dimensional $(A+R+G+B)$ video matrices and sound
signals in order to apply some well-known alchemical techniques to
the video at real-time to get a mirror effect and accompanying
transmutation and transformation stages of the video based on the
stability of the sound produced for some duration of time.


\section{Brief Review of References Used}
\index{Introduction!Review of References}

A number of alchemical works have been used as a reference
material to create the new mirror-based process of purification.
Additionally, a modern unusual physics-based dynamics book~\cite{landau-comp-physics-97}
was used to create and test some expressions in the Jitter
environment. A number of reference include the Jitter and Max/MSP
tutorial patches for motion tracking, finding bounds, extracting
FFT off a video matrix, and getting pitch (FFT-based) and an
amplitude information from an external \api{pitch\~} patch
done by Tristan Jehan~\cite{external-jitter-pitch}.

The two previous works by Eidelman~\cite{lizets-heart} and
Mokhov~\cite{mokhov-quintessence} were taken as the initial foundation in this
collection of Jitter patches; while a lot of those unused in this
particular projects, they are still found the \file{patches} directory
of this work.
An accompanying QuickTime movie \file{alchymical-mirror.mov} gives
an approximate summary how the process works. Some concepts and ideas from other, less related work,
had also some impact in here, e.g.
from~\cite{edmund-bacon,mackay-delusions-crowds,mackay-delusions-memoirs,fairweather-basil}.


	\chapter{Methodology and Implementation}
\index{Methodology}
\index{Implementation}

The origins of this work in portions resulted from the work
of extracting a quintessence of gold by Agricola
in \cite{agricola-gold} and perfection of the alchemists
themselves while in the process
study done by Newman in \cite{newman-ambitions}.
Like gold, a soul may have a various degree
of purity. A common golden or red color of 
soul represents most pure, so while traveling
through the stages of perfection, and alchemists
should observe some reddening of the source video
stream at all stages but the last.
The implementation consists of a set of \file{alchemy.*} patches.
The starting-point patch for this work is \api{alchemy.tracking.alpha}.

\section{Master Patch -- \api{alchemy.tracking.alpha}}

Our master patch summons all the sub-patches and relates them one to the other.
It begins by taking the streaming movie from the \api{camerainput} patch of Freida Abtan and
send it to the \api{alchemy.mirror} one that reflects it by the $y$ axis, giving
the mirror effect.  It also triggers the \api{alchemy.pitch} sub-patch
that turns on the analysis of the sound coming into the microphone.  It analyzes
the sound and depending on its duration and stability varies the performer's
level of progression and gives out the progression percentage to the next
level.  This percentage is demonstrated by the value of the $R$ channel in the
first 3 levels of the experience.  Depending on the level of progression,
other sub-patches are triggered, helping the user to visualize his/her progress
level.  The sub-patches are triggered in the following order:

\begin{itemize}
\item
level 1: direct movie from the \api{alchemy.mirror} object;
\item
level 2: \api{alchemy.mirror.bugs};
\item
level 3: \api{alchemy.dissolver};
\item
level 4(final): \api{alchemy.water.solvent} that previously passed by 
the \api{alchemy.mirror.star} object.
Here the star controlled by the \api{alchemy.findbounds} object
was added to the original mirrored movie.  Reaching the level 4,
the user is able to control the star using his hands or other
objects of a color set in the \api{alchemy.findbounds} patch.
\end{itemize}

The follow up sections describe more the patches mentioned above.

\section{Dissolving Colors with \api{alchemy.water.solvent}}

This patch piece has already made appearance in the Alchemy Framework
in \cite{mokhov-quintessence}. Here its description is summarized again
for convenience.

As it has been seen in the summary of alchemical works \cite{humburg-colors}
done by various groups, there is a set of colors attribute
to the transformation of any type of matter.
The color green was often attributed to the allegorical
{\em Green Lion} representing the perfect gold. The color
red was a common attribute of the gold at the final stage
of preparation (resurrection after being burnt).
As it's been known for ages,
the two colors combined give the color yellow, a day-to-day
visual attribute of gold.

Thus dissolving our source material of video into basic $R$,
$G$, and $B$ planes, and combining $R$ and $G$ is the first step
in the process. The color blue ($B$) is discarded.
Then the original $(A+R+G+B)$ material and the extracted pre-golden
material $(A+R+G)$ are used further in the process.

Through so-called solvent water patch, \api{alchemy.water.solvent},
the visual media stream is split into the $R, G, B$ components out of
which only $R,G$ along with alpha $A$ are retained, as described
earlier. The patch
simply unpacks the planes with \api{jit.unpack} and then recombines the needed ones
with \api{jit.pack}.
Thus, the output has three matrices: $(A+R+G)$, $(A+G)$, and $(A+R)$.
In the \api{alchemy.tracking.alpha} patch the last two are unused and the result
is only achievable of after passing through the other two stages representing
purification through ``burning'' with \api{alchemy.mirror.bugs} and ``calcination''
with \api{alchemy.dissolver}, though their names do not correspond exactly
to what those patches do as of this writing :-).


\section{Mirroring -- \api{alchemy.mirror}}

This patch takes a movie matrix and reflects it by the $y$ axis, resulting in a real
mirroring effect.
The mirroring patch takes as input a streaming movie (in our case the movie
is coming from the \api{camerainput} patch).  Using the \api{jit.mxform2d} 
object, which applies linear algebra operations on matrices, the movie
is mirrored by the $y$ axis and the final result is outputted 
in the only outlet.

\section{Finding Bounds -- \api{alchemy.findbounds}}

This patch is used to find the outline box of a user-specified color and output
center coordinates of the box, its height and width.
The \api{alchemy.findbounds} patch takes a streaming movie in the matrix form
as its input and sends it 
to a \api{jit.pwindow}.  A \api{suckah} object placed right on top of the movie matrix permits 
quick access to the its colors.  By clicking
on the movie, we can define which color must be identified; we also allow a small
variation in the color that could result due to the lighting of the room where the
experiment takes place.  The minimum and the maximum color values, as well as
the input movie matrix, are sent to the \api{jit.findbounds} object that outputs
the bounding coordinates of the color.  Taking those coordinates, the \api{jit.lcd}
object draws an outline box of the color.  We also use those coordinates to compute
and output in the four outlets the center coordinates $(x,y)$ of the outline box
 and it's width and height.

\section{Finding Pitch and an Apmplitude -- \api{alchemy.pitch}}

This patch is used to take a sound input and depending on the pitch, amplitude
and duration of the sound, outputs the level achieved by the user and the percentage
of the user's progression to the next level.

As the patch is triggered, the microphone is turned on and the sound is directed
for analysis into the \api{pitch\~} object that returns an amplitude and pitch
values of the audio signal. Amplitude and Pitch analysis that follows are
exactly similar so we will describe only one procedure as the other one is analogous.
The values are analyzed at a $200 ms$ interval because natural human voice
changes gradually if analyzed at very small deltas $\Delta{f}$ and $\Delta{a}$
as we allow a small fluctuation in the voice, if it would be analyzed continuously,
any sound would be qualified as homogeneous.  So we decided to analyze it ever
roughly 5 times per minute.  The value is then compared to the previous one to
see the fluctuation.  If it is in an acceptable range, then the value is passed 
on and the timer continues to counts the duration of that homogeneous sound.
Otherwise, if the fluctuation is too big, a bang is generated signifying that
the timer should restart since the sound has changed.
The timer displays the duration of the sound.  If the duration was long
enough for a certain level, a bang is sent to the counter saying that the level must be changed.
We have generated many outlets for debugging purposes, however the only two
outlets that were used by the final program were the first one, giving the
percentage of completion of the current level and the second one,
displaying the level reached at present point.

\section{Dissolver Expressions -- \api{alchemy.dissolver}}

The \api{alchemy.dissolver} patch has changed since the last version of
the Alchemy Framework presented in \cite{mokhov-quintessence}. It does no longer
do very basic frame-difference-based motion tracking with the matrix-wide
operator \texttt{!-}. It accepts an additional input parameter and has a
few expressions implemented based on the Chapter 13 of \cite{landau-comp-physics-97}.
Out of the four expressions only a combination of the 2nd and 3rd is
activated to simulate ``calcination-in-motion''. The 2nd inlet to the patch
now is the parameter to those expressions. This patch may get renamed at
a later point.

\section{\api{alchemy.mirror.bugs}}

This not-very-well-named patch acts as a source of ``purification-though-burning''
of an image with a noise and a projected goal of the star in it as an ultimate
goal. There are two parts to this patch. The first one applies an expression
on input video matrix from the random numbers logistic map adapted equation
from \cite{landau-comp-physics-97} where the parameter $\$i$ may vary from
the progress reported by the \api{alchemy.pitch}. This piece is matrix-subtracted
from a fractal noise generated by the \api{jit.bfg} object used in its example.
The two are averaged and the FFT is extracted to form a star in the middle
(basically FFT repeated 4 times in 4 quadrants in the matrix) and then all
these pieces (the transmuted video, the fractal noise, and the FFT star)
are combined together at the output.
(The name \texttt{.bugs} was used to indicate that there were many bugs
with expression previously, before it ended up in the final patch.)

\section{\api{alchemy.mirror.star}}

This is the final stage patch that takes effect when the final
``golden'' stage is reached. In that stage the output of the \api{alchemy.findbounds}
patch goes as an input to this patch along with the output of \api{alchemy.water.solvent}.
This includes the coordinates of the center point of the bounding box (used to place
the star) and the height and the width of the box. The height and the width are used
to properly scale the coordinates and the star size so when supposedly one
moves their hands (with distinctly colored gloves) apart it should grow bigger, and if one places their hand
closer together it should be smaller. Likewise, if the hands with the This effect works, but is quite unstable
and requires more calibration. The star itself is a FFT-quadruple of the video
signal coming it, and as such, is shining-animated.

First, the FFT transform of the video matrix is taken to determine the star's shape.
Then, the star is scaled based on the height/width ration we get as an input
using \api{jit.mxform2d}. Further, the star is repositioned using \api{jit.repos}
and the $x$ and $y$ offsets calculated from the center coordinates received.
Finally, this altered star is combined with the original incoming video matrix
to produce the final result.

\section{Recording -- \api{alchemy.qt.record}}

This is a wrapper around the \api{jit.qt.record} object.  We changed it to
start recording the movie as soon as our patch begins, to automatically
record it as real-time video and we added another inlet to be able to
stop the recording whenever needed.


	\chapter{Conclusions}
\index{Conclusion}

\section{Limitations}
\index{Limitations}

The most prominent limitation of the patch as of this writing
is instability in extracting and catching and navigating the
star at the final stage.

\section{Acknowledgments}
\index{Acknowledgments}

\begin{itemize}
\item
Dr. Xin Wei Sha, for unconventional course on Alchemy,
Real-time Media, and Calligraphic Video and allowing us
using the TML resources.
\item
Freida Abtan, for an awesome Jitter intro.
\end{itemize}

\section{Future Work}
\index{Future Work}

First, the future work will focus on stabilizing the star-bounds
effect.
Additionally, a supplemental degree of sound feedback will
be added as it was not at all addressed yet in the
present work. Finally, more experimentation with physics
and the corresponding expressions as well as attaining more
realistic mirror effects, and maybe an installation alongside
with the TML.


	\addcontentsline{toc}{chapter}{Bibliography}

\bibliographystyle{alpha}
\bibliography{alchymical-mirror-hci-jitter}



	\printindex
\end{document}